\documentclass[notitlepage,letterpaper,11pt]{article}

\usepackage[left=1in, width=6.75in]{geometry}

\usepackage{times}

\usepackage{setspace}
\usepackage{color}
\usepackage{amsmath}
\usepackage{amsfonts}
\usepackage{amssymb}
\usepackage{graphicx}
\usepackage{booktabs}
\usepackage{multirow}
\usepackage{hyperref}

\title{Computational fact checking from knowledge networks}

\author{
    Giovanni Luca Ciampaglia\textsuperscript{1}\thanks{To whom correspondence should be addressed; E-mail: gciampag@indiana.edu.},
    Prashant Shiralkar\textsuperscript{1},
    Luis M. Rocha\textsuperscript{1,2},\\
    Johan Bollen\textsuperscript{1},
    Filippo Menczer\textsuperscript{1},
    Alessandro Flammini\textsuperscript{1}\\
    \normalsize{\textsuperscript{1}Center for Complex Networks and Systems Research,}\\
    \normalsize{School of Computing, Indiana University,}\\
    \normalsize{Bloomington, IN USA}\\
    \normalsize{\textsuperscript{2}Instituto Gulbenkian de Ciencia,}\\
    \normalsize{Oeiras, Portugal}
}

\date{}

\begin{document}

\renewcommand{\thefootnote}{\fnsymbol{footnote}}
\maketitle
\renewcommand{\thefootnote}{\arabic{footnote}}

\doublespacing
\raggedright


\begin{abstract}

Traditional fact checking by expert journalists cannot keep up with the enormous
volume of information that is now generated online. Computational fact checking
may significantly enhance our ability to evaluate the veracity of dubious
information. Here we show that the complexities of human fact checking can be
approximated quite well by finding the shortest path between concept nodes under
properly defined semantic proximity metrics on knowledge graphs. Framed as a
network problem this approach is feasible with efficient computational
techniques. We evaluate this approach by examining tens of thousands of claims
related to history, entertainment, geography, and biographical information using
a public knowledge graph extracted from Wikipedia. Statements independently
known to be true consistently receive higher support via our method than do
false ones. These findings represent a significant step toward scalable
computational fact-checking methods that may one day mitigate the spread of
harmful misinformation.

\end{abstract}

\section*{Introduction}

Online communication platforms, in particular social media, have created a
situation in which the proverbial lie ``can travel the world before the truth
can get its boots on.'' Misinformation \cite{Mendoza2010}, astroturf
\cite{Ratkiewicz2011}, spam \cite{Cranor1998}, and outright fraud
\cite{Jagatic2007} have become widespread. They are now seemingly unavoidable
components of our online information ecology \cite{Friggeri2014} that jeopardize
our ability as a society to make rapid and informed decisions
\cite{Flanagin2000,Rieh2007,Kata2010,Castillo2011,Lewandowsky2012}.

While attempts to partially automate the detection of various forms of
misinformation are burgeoning
\cite{CJR-factchecking,Gupta2014,Resnick2014,Wu2014,Finn2014}, automated reasoning
methods are hampered by the inherent ambiguity of language and by deliberate
deception. However, under certain conditions, reliable knowledge transmission
can take place online \cite{Berners-Lee2001}. For example, Wikipedia, the
crowd-sourced online encyclopedia, has been shown to be nearly as reliable as
traditional encyclopedias, even though it covers many more topics \cite{Giles2005}.
It now serves as a large-scale knowledge repository for millions of individuals,
who can also contribute to its content in an open way. Vandalism, bias,
distortions, and outright lies are frequently repaired in a matter of minutes
\cite{Priedhorsky2007}. Its continuous editing process even indicates signs of
collective human intelligence \cite{Dedeo2013}.

Here we show that we can leverage any collection of factual human knowledge,
such as Wikipedia, for automatic fact checking \cite{Cohen2011}. Loosely
inspired by the principle of epistemic closure \cite{Luper2012}, we
computationally gauge the support for statements by mining the connectivity
patterns on a knowledge graph. Our initial focus is on computing the support of
simple statements of fact using a large-scale knowledge graph obtained from
Wikipedia.

\subsection*{Knowledge Graphs} 

Let a \emph{statement of fact} be represented by a subject-predicate-object
triple, e.g., (``Socrates,'' ``is a,'' ``person''). A set of such triples can be
combined to produce a \emph{knowledge graph} (KG), where nodes denote
\emph{entities} (i.e., subjects or objects of statements), and edges denote
predicates. Given a set of statements that has been extracted from a knowledge
repository --- such as the aforementioned Wikipedia --- the resulting KG network
represents all factual relations among entities mentioned in those statements.
Given a new statement, we expect it to be true if it exists as an edge of the
KG, or if there is a short path linking its subject to its object within the KG.
If, however, the statement is untrue, there should be neither edges nor short
paths that connect subject and object.

In a KG distinct paths between the same subject and object typically provide
different factual support for the statement those nodes represent, even if the
paths contain the same number of intermediate nodes. For example, paths that
contain generic entities, such as ``United States'' or ``Male,'' provide weaker
support because these nodes link to many entities and thus yield little specific
information. Conversely, paths comprised of very specific entities, such as
``positronic flux capacitor'' or ``terminal deoxynucleotidyl transferase,''
provide stronger support. A fundamental insight that underpins our approach is
that the definition of path length used for fact checking should account for
such information-theoretic considerations. 

\begin{figure}
    \centering
    \includegraphics{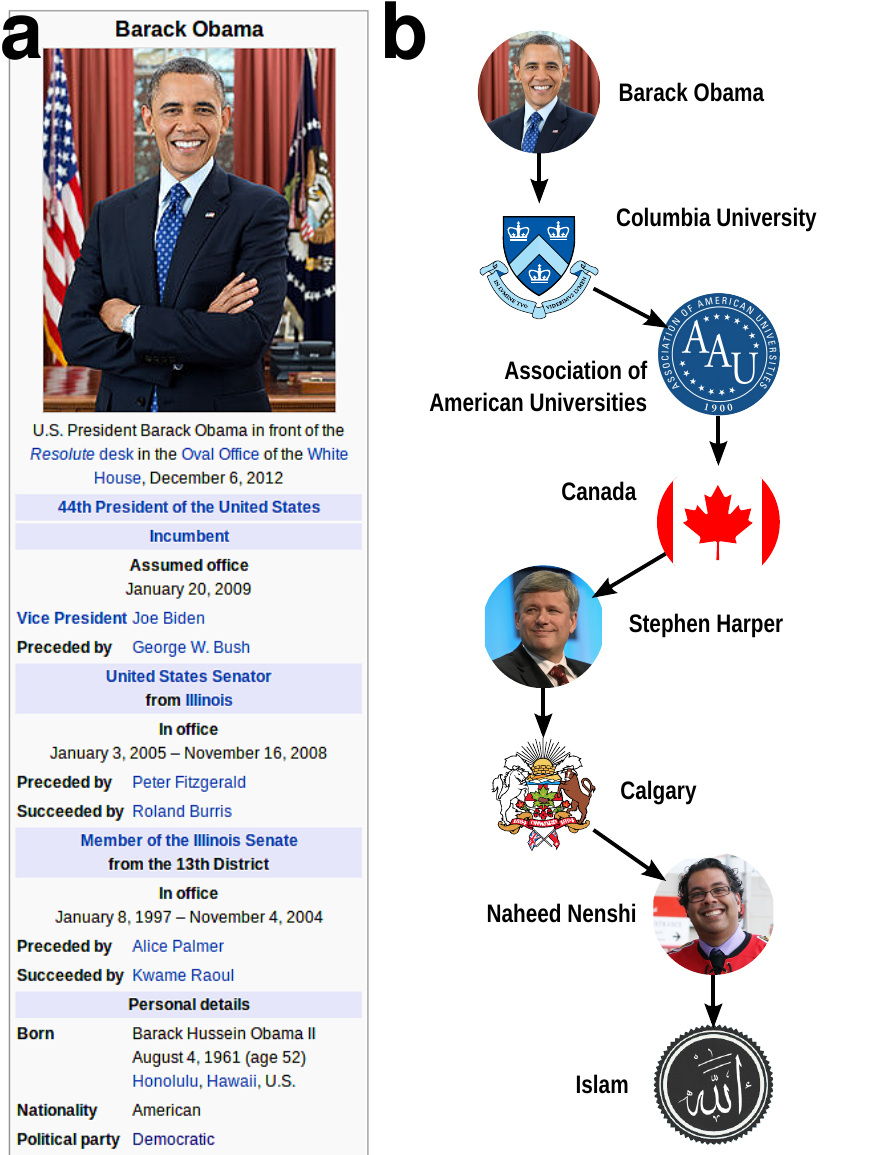}
    \caption{\textbf{Using Wikipedia to fact-check statements}. \textbf{(a)} To
    populate the knowledge graph with facts we use structured information contained
    in the `infoboxes' of Wikipedia articles (in the figure, the infobox of the
    article about \emph{Barack Obama}). \textbf{(b)} Using the Wikipedia
    Knowledge Graph, computing the truth value of a subject-predicate-object
    statement amounts to finding a path between subject and object entities. In
    the diagram we plot the shortest path returned by our method for the
    statement ``\emph{Barack Obama} is a \emph{muslim}.'' The path traverses
    high-degree nodes representing generic entities, such as
    \emph{Canada}, and is assigned a low truth value.}
\label{fig:fig1_diagram}
\end{figure}

To test our method we use the DBpedia database \cite{Auer2007}, which consists
of all factual statements extracted from Wikipedia's ``infoboxes'' (see
Fig.~\ref{fig:fig1_diagram}(a)). From this data we build the large-scale
\emph{Wikipedia Knowledge Graph} (WKG), with 3 million entity nodes linked by
approximately 23 million edges (see \emph{Materials and Methods}). Since we use
only facts within infoboxes, the WKG contains the most uncontroversial
information available on Wikipedia. This conservative approach is employed to
ensure that our process relies as much as possible on a human-annotated,
collectively-vetted factual basis. The WKG could be augmented with automatic
methods to infer facts from text and other unstructured sources available
online. Indeed, other teams have proposed methods to infer knowledge from text
\cite{Dong2014} to be employed in large and sophisticated rule-based inference
models \cite{Etzioni2008,Niu2012,Dong2014}. Here we focus on the feasibility of
automatic fact checking using \emph{simple} network models that leverage
DBpedia. For this initial goal, we do not need to enhance the WKG, but such
improvements can later be incorporated.

\subsection*{Semantic Proximity from Transitive Closure}

Let the WKG be an undirected graph $G = \left( V,E \right)$ where $V$ is a set
of concept nodes and $E$ is a set of predicate edges (see \emph{Materials and
Methods}). Two nodes $v, w \in V$ are said to be \emph{adjacent} if there is an
edge between them $(v, w) \in E$. They are said to be \emph{connected} if there
a sequence of $n \ge 2$ nodes $v = v_1, v_2, \dots v_n = w$, such that, for
$i=1,\dots, n - 1$ the nodes $v_i$ and $v_{i+1}$ are adjacent. The
\emph{transitive closure} of $G$ is $G^\ast = \left( V, E^{\ast} \right)$ where
the set of edges is closed under adjacency, that is, two nodes are adjacent in
$G^\ast$ \emph{iff} they are connected in $G$ via at least one path. This
standard notion of closure has been extended to weighted graphs, allowing
adjacency to be generalized by measures of path length \cite{Simas2014}, such as
the semantic proximity for the WKG we introduce next.

The truth value $\tau(e) \in [\, 0, 1 \,]$ of a new statement $e = (s, p, o)$ is
derived from a transitive closure of the WKG. More
specifically, the truth value is obtained via a path evaluation function:
$\tau(e) = \max \, \mathcal{W}(P_{s,o})$. This function maps the set of possible
paths connecting $s$ and $o$ to a truth value $\tau$. A path has the form
$P_{s,o} = v_1 v_2 \ldots v_n$, where $v_i$ is an entity node, $( v_i, v_{i+1}
)$ is a edge, $n$ is the path length measured by the number of its constituent
nodes, $v_1 = s$, and $v_n = o$. Various characteristics of a path can be taken
as evidence in support of the truth value of $e$. Here we use the
\emph{generality} of the entities along a path as a measure of its length, which
is in turn aggregated to define a \emph{semantic proximity}:
\begin{equation}
\mathcal{W}(P_{s,o}) = \mathcal{W}(v_1 \ldots v_n) = \left[1 + \sum_{i = 2}^{n - 1} \log k\left( v_i \right)\right]^{-1}
\label{eq:evalfunction}
\end{equation}
where $k( v )$ is the degree of entity $v$, i.e., the number of WKG statements
in which it participates; it therefore measures the generality of an entity. If
$e$ is already present in the WKG (i.e., there is an edge between ~$s$ and $o$),
it should obviously be assigned maximum truth. In fact $\mathcal{W} = 1$ when $n
= 2$ because there are no intermediate nodes. Otherwise an indirect path of
length $n > 2$ may be found via other nodes. The truth value $\tau( e )$
maximizes the semantic proximity defined by Eq.~\ref{eq:evalfunction}, which is
equivalent to finding the shortest path between $s$ and $o$ \cite{Simas2014}, or
the one that provides the maximum information content
\cite{Markines2009,Aiello2011Friendship} in the WKG. The transitive closure of
weighted graphs equivalent to finding the shortest paths between every pair of
nodes is also known as the \emph{metric closure} \cite{Simas2014}.
Fig.~\ref{fig:fig1_diagram}(b) depicts an example of a shortest path on the WKG
for a statement that yields a low truth value.

Note that in this specific formulation we disregard the semantics of the
predicate, therefore we are only able to check statements with the simplest
predicates, such as ``is a''; negation, for instance, would require a more
sophisticated definition of path length.

Alternative definitions of $\tau( e )$ are of course possible. Instead of
shortest paths, one could use a different optimization principle, such as widest
bottleneck, also known as the \emph{ultra-metric closure} \cite{Simas2014},
which corresponds to maximizing the path evaluation function $\mathcal W_u$:
\begin{align}
    \mathcal{W}_u(P_{s,o}) = & \, \mathcal{W}_u\left(v_1 \ldots v_n\right) =
    \nonumber \\ = & \left\{
 \begin{array}{c l}
     1   &   n = 2\\
     \left[1 + \max_{i = 2}^{n - 1} \left\{\log k\left( v_i \right)\right\} \right]^{-1} & n>2.
 \end{array}
 \right.
\label{eq:evalultrafunction}
\end{align}
Or it could be possible to retain the original directionality of
edges and have a directed WKG instead of an undirected one.
As described next, we evaluated alternative definitions of $\tau( e )$
and found Eq.~\ref{eq:evalfunction} to perform best.

\section*{Results}

\subsection*{Calibration} \label{sec:calibration} 

\begin{figure}[b!]
    \centering
    \includegraphics[width=0.49\textwidth]{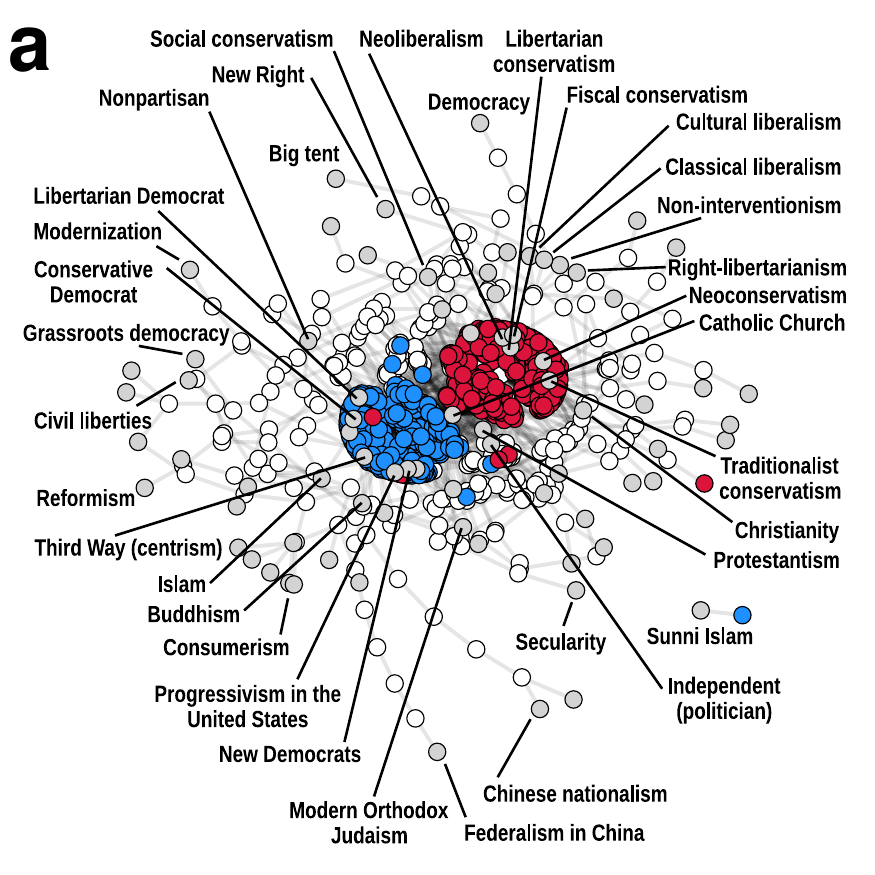}
    \includegraphics[width=0.49\textwidth]{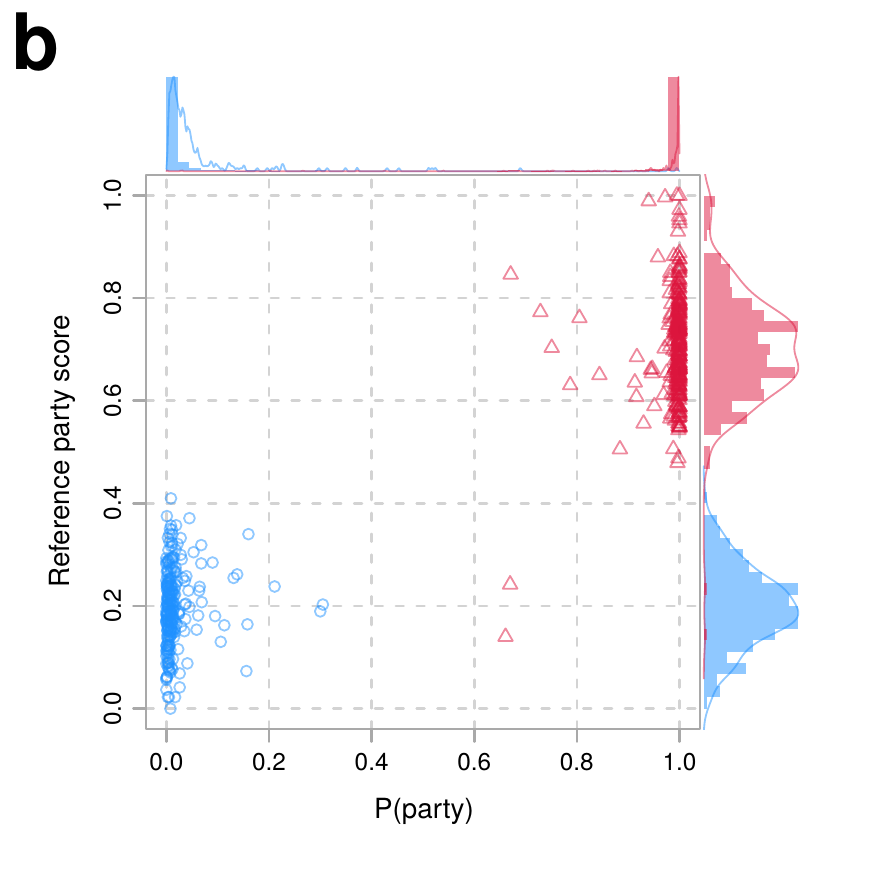}
    \caption{\textbf{Ideological classification of the US Congress based on
    truth values}. \textbf{(a)} Ideological network of the 112th US
    Congress. The plot shows a subset of the WKG constituted by paths between
    Democratic or Republican members of the 112th US Congress and various
    ideologies. Red and blue nodes correspond to members of Congress, gray nodes
    to ideologies, and white nodes to vertices of any other type. The position
    of the nodes is computed using a force-directed layout \cite{Kamada1989},
    which minimizes the distance between nodes connected by an edge weighted by
    a higher truth value. For clarity only the most significant paths,
    whose values rank in the top 1\% of truth values,
    are shown.
    \textbf{(b)} Ideological classification of members of the 112th US Congress.
    The plot shows on the $x$ axis the party label probability given by a Random
    Forest classification model trained on the truth values computed on the WKG,
    and on the $y$ axis the reference score provided by \textsc{dw-nominate}.
    Red triangles are members of Congress affiliated to the Republican party and
    blue circles to the Democratic party. Histograms and density estimates of
    the two marginal distributions, color-coded by actual
affiliation, are shown on the top and right axes.}
\label{fig:figS1_classification} \end{figure}

Our fact-checking method requires that we define a measure of path semantic
proximity by selecting a transitive closure algorithm (the shortest paths of
Eq.~\ref{eq:evalfunction} or the widest bottleneck paths of
Eq.~\ref{eq:evalultrafunction}) and a directed or undirected WKG representation.
To evaluate these four combinations empirically, let us attempt to infer
the party affiliation of US Congress members. In other words, we want to
compute the support of statements like ``$x$ is a member of $y$'' where $x$ is a
member of Congress and $y$ is a political party. We consider all members of the
112th US Congress that are affiliated with either the Democratic or Republican
party (Senate: $N = 100$; House: $N = 445$). We characterize each member of
Congress with its semantic proximity to all nodes in the WKG that represent
ideologies. This yields an $N \times M$ feature matrix $\mathcal F_\mathrm{tc}$
for each of the four transitive closure methods. The top panel of
Fig.~\ref{fig:figS1_classification} illustrates the proximity network obtained
from $\mathcal F_\mathrm{tc}$ that connects members of the 112th Congress and
their closest ideologies, as computed using Eq.~\ref{eq:evalfunction}. A high
degree of ideological polarization can be observed in the WKG, consistent with
blogs \cite{Adamic2005} and social media \cite{Conover2011}.

We feed $\mathcal F_\mathrm{tc}$ into off-the-shelf classifiers (see
\emph{Materials and Methods}). As shown in Table~\ref{tab:calibrationresults},
the metric closure on the undirected graph gives the most accurate results.
Therefore, we continue to use this combination in our semantic proximity
computations when performing the validation tasks described below.

To evaluate the overall performance of the calibrated model, we also compared it
against \textsc{dw-nominate}, the state of the art in political classification
\cite{Poole2007}. This model is not based on data from a knowledge graph, but on
explicit information about roll-call voting patterns. Comparing our
classification results with such a baseline is also useful to gauge the quality
of the latent information contained in the WKG for the task of political
classification. As shown in the bottom panel of
Fig.~\ref{fig:figS1_classification}, a Random Forests classifier trained on our
truth values matches the performance of \textsc{dw-nominate}.

\subsection*{Graph search vs infoboxes only} 

Most of the WKG information that our fact checker exploits is provided by
indirect paths (i.e., comprising $n>2$ nodes). To demonstrate this, we compare
the calibrated model of Eq.~\ref{eq:evalfunction} to the fact checker's
performance with only the information in the infoboxes.

In practice, we compute an additional feature matrix $\mathcal F_\mathrm{b}$,
using the same sequence of steps outlined in the calibration phase, but
additionally constraining the shortest path algorithm to use only paths (if any)
with exactly $n = 2$ nodes, i.e., direct edges. Thus $\mathcal F_\mathrm{b}$
encodes only the information of the infoboxes of the politicians. The results
from 10-fold cross validation using $\mathcal F_\mathrm{tc}$ and $\mathcal
F_\mathrm{b}$ are shown in Table~\ref{tab:classification}. The same
off-the-shelf classifiers, this time trained on $\mathcal F_\mathrm{b}$, perform
only slightly better than random, thus confirming that the truth signal is
yielded by the structure of indirect connections in the WKG.

\subsection*{Validation on factual statements} 

\begin{figure}
    \centering
    \includegraphics[width=.94\textwidth]{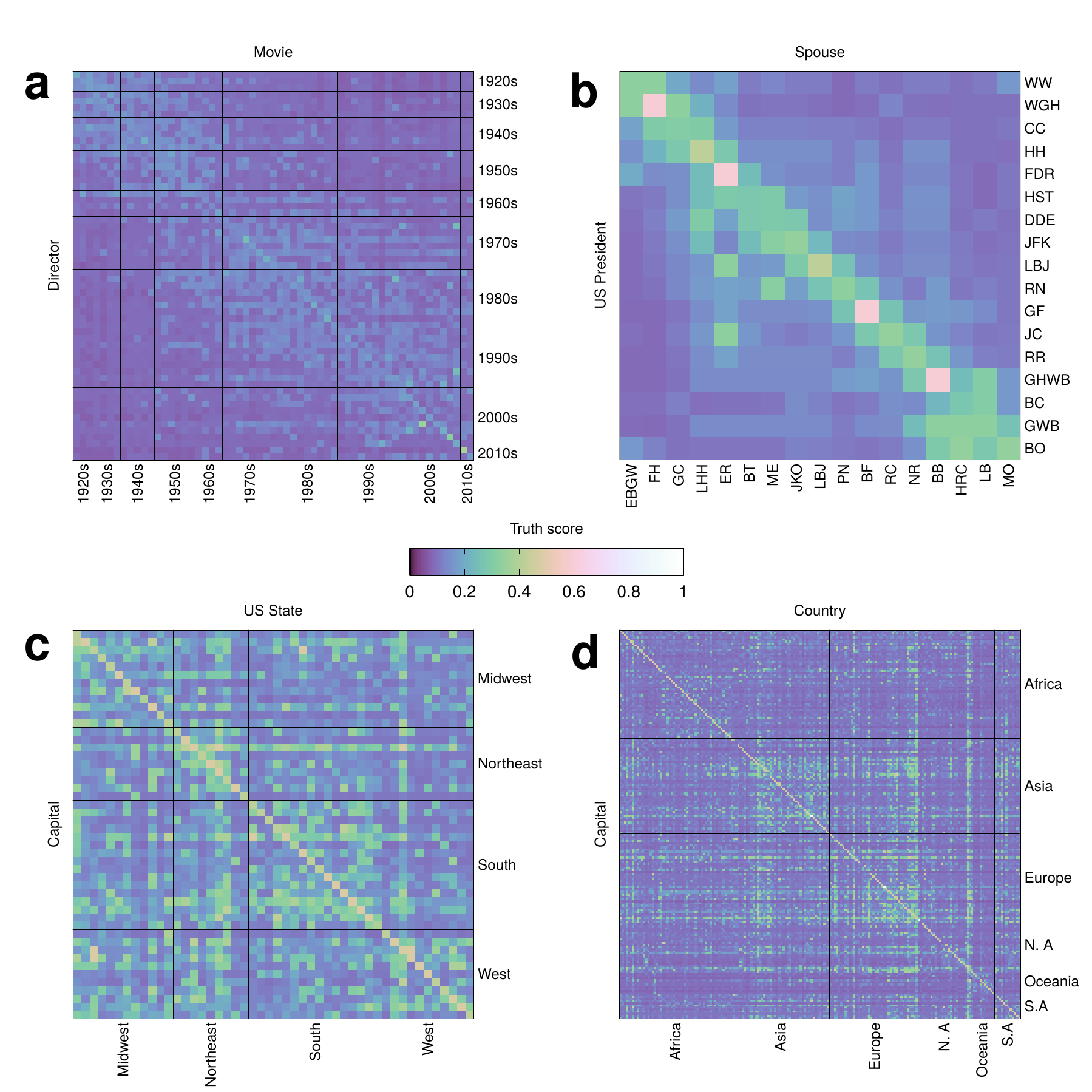}
    \caption{\textbf{Automatic truth assessments for simple factual statements.}
    In each confusion matrix, rows represent subjects and columns represent objects.
    The diagonals represent true statements. Higher truth values are mapped to
    colors of increasing intensity. \textbf{(a)} Films winning the Oscar
    for Best Movie and their directors, grouped by decade of award
    (see the complete list in the \emph{SI appendix}).
    \textbf{(b)} US presidents and their spouses, denoted by
    initials. \textbf{(c)} US states and their capitals, grouped by US
    Census Bureau-designated regions. \textbf{(d)} World countries
    and their capitals, grouped by continent.}
\label{fig:fig2_confusion}
\end{figure}

\begin{figure}[htbp]
  \centering
  \includegraphics{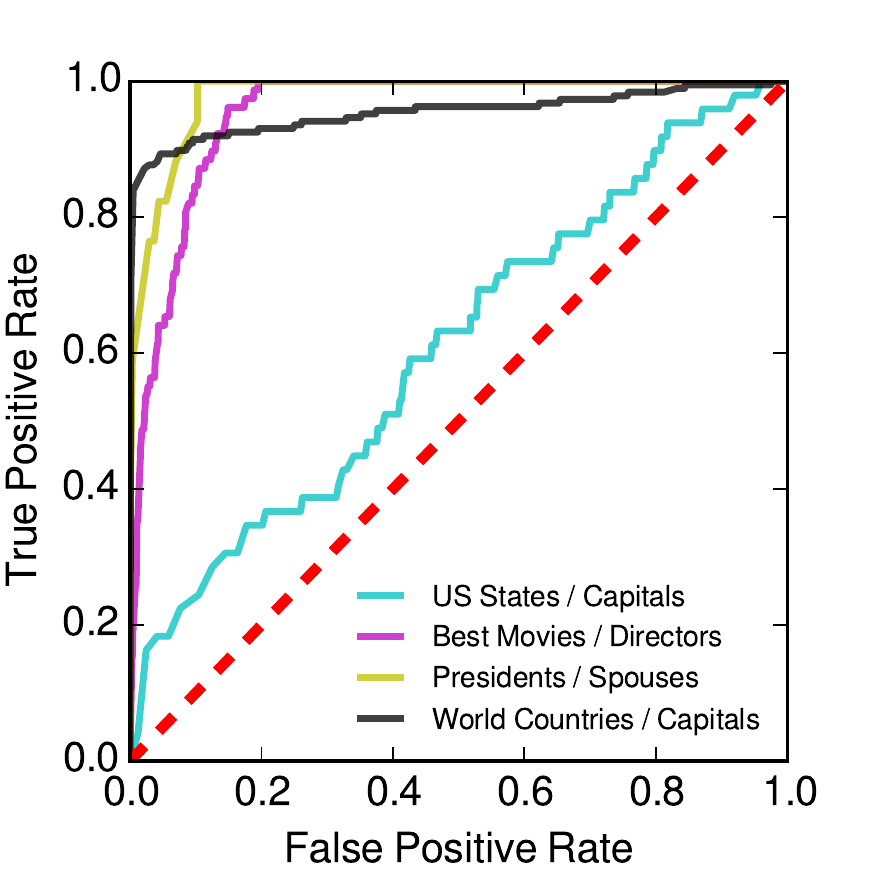}
  \caption{\textbf{Receiver Operating Characteristic for the multiple questions
  task}. For each confusion matrix depicted in Fig.~\ref{fig:fig2_confusion}
  we compute ROC curves where true statements correspond to the
diagonal and false statements to off-diagonal elements. The red dashed line
represents the performance of a random classifier.}
  \label{fig:figS2_multiple_questions_roc}
\end{figure}

We test our fact-checking method on tasks of increasing difficulty, and begin by
considering simple factual statements in four subject areas related to
entertainment, history, and geography. We evaluate statements of the form
``$d_i$ directed $m_j$,'' ``$p_i$ was married to $s_j$,'' and ``$c_i$ is the
capital of $r_j$,'' where $d_i$ is a director, $m_j$ is a movie, $p_i$ is a US
president, $s_j$ is the spouse of a US president, $c_i$ is a city, and $r_j$ is
a country or US state. By considering all combinations of subjects and objects
in these classes, we obtain matrices of statements (see \emph{Materials and
Methods}). Many of them, such as ``Rome is the capital of India,'' are false.
Others, such as ``Rome is the capital of Italy,'' are true. To prevent the task
from being trivially easy, we remove any edges that represent true statements in
our test set from the graph. Fig.~\ref{fig:fig2_confusion} shows the matrices
obtained by running the fact checker on the factual statements. Let $e$ and $e'$
be a true and false statement, respectively, from any of the four subject areas.
To show that our fact checker is able to correctly discriminate between true and
false statements with high accuracy, we estimate the probability that
$\tau\left( e \right) > \tau\left( e' \right)$. To do so we plot the ROC curve
of the classifier (see Fig.~\ref{fig:figS2_multiple_questions_roc}) since the
area under the ROC curve is equivalent to this probability \cite{Fawcett2006}.
With this method we estimate that, in the four subject areas, true statements
are assigned higher truth values than false ones with probability 95\%, 98\%,
61\%, and 95\%, respectively.

\subsection*{Validation on annotated corpus} 

\begin{figure}
    \centering
    \includegraphics{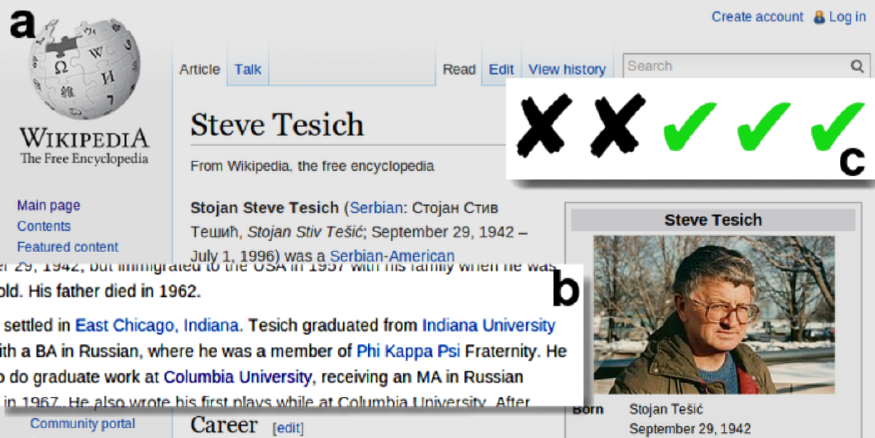}
    \includegraphics{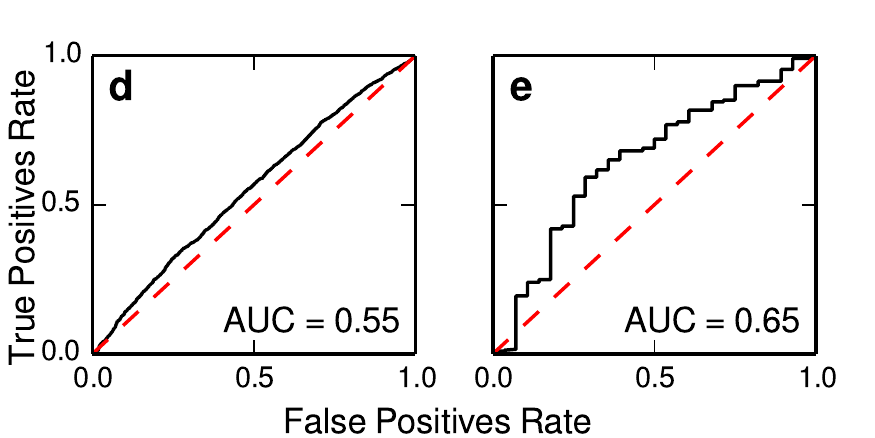}
    \caption{\textbf{Real-world fact-checking scenario}. \textbf{(a)}~A document
    from the ground truth corpus. \textbf{(b)}~Statement to fact-check:
    \textit{Did Steve Tesich graduate from Indiana University, Bloomington?} This
    information is not present in the infobox, and thus it is not part of the
    WKG. \textbf{(c)}~Annotations from five human raters. In this case, the
    majority of raters believe that the statement is true, and thus we consider
    it as such for classification purposes. \textbf{(d)}~Receiver operating
    characteristic (ROC) curve of the classification for
    subject-predicate-object statements in which the predicate is
    ``institution'' (e.g., ``Albert Einstein,'' ``institution,'' ``Institute for
    Advanced Studies''). A true positive rate above the false positive rate
    (dashed line), and correspondingly an area under the curve (AUC) above 0.5,
    indicate better than random performance. \textbf{(e)}~ROC curve for
    statements with ``degree'' predicate (e.g., ``Albert Einstein,''
    ``degree,'' ``University Diploma'').}
    \label{fig:fig3_relation}
\end{figure}

In a second task, we consider an independent corpus of novel statements
extracted from the free text of Wikipedia and annotated as true or false by
human raters \cite{Google2013} (see \emph{Materials and Methods}). We compare
the human ratings with the truth values provided by our automatic fact checker
(Fig.~\ref{fig:fig3_relation}). Although the statements under examination
originate from Wikipedia, they are not usually represented in the WKG, which is
derived from the infoboxes only. When a statement is present in the WKG, the
link is removed. The information available in the WKG about the entities
involved in these particular statements is very sparse, therefore this task is
more difficult than the previous case.

We find that the truth values computed by the fact checker are
positively correlated to the average ratings given by the human evaluators.
Table~\ref{tab:relation} shows the positive correlation between GREC human
annotations and our computational truth scores.

As shown in Fig.~\ref{fig:fig3_relation}, our fact checker yields consistently
higher support for true statements than false ones. Using only information in
the infoboxes however yields worse performance, closer to random choice:
$\mathrm{AUROC} = 0.47$ and $0.52$ for the `degree' and `institution'
predicates, respectively. We conclude that the fact checker is able to integrate
the strength of indirect paths in the WKG, which pertain to factual information
not originally included in the infoboxes.

\section*{Discussion}

These results are both encouraging and exciting: a simple shortest path computation
maximizing information content can leverage an existing body of collective human
knowledge to assess the truth of new statements. In other words, the important
and complex human task of fact checking can be effectively reduced to a simple
network analysis problem, which is easy to solve computationally. Our approach
exploits implicit information from the topology of the WKG, which is different
from the statements explicitly contained in the infoboxes. Indeed, if we base
our assessment only on direct edges in the WKG, performance decreases significantly. This
demonstrates that much of the correct measurement of the truthfulness of
statements relies on indirect paths. Because there are many ways to compute
shortest paths in distance graphs, or transitive closures in weighted graphs
\cite{Simas2014}, there is ample room for improvement on this method.

We live in an age of overabundant and ever-growing information, but much of it
is of questionable veracity \cite{Lewandowsky2012,Nyhan2013}. Establishing the
reliability of information in such circumstances is a daunting but critical
challenge. Our results show that network analytics methods, in conjunction with
large-scale knowledge repositories, offer an exciting new opportunity towards
automatic fact-checking methods. As the importance of the Internet in our
everyday lives grows, misinformation such as panic-inducing rumors, urban
legends, and conspiracy theories can efficiently spread online in variety of new
ways \cite{Kata2010,Friggeri2014}. Scalable computational methods, such as the
one we demonstrate here, may hold the key to mitigate the societal effects of
these novel forms of misinformation. 

\section*{Materials and Methods}

\subsection*{Wikipedia Knowledge Graph} %
To obtain the WKG we downloaded and parsed RDF triples data from the DBpedia
project (\url{dbpedia.org}). We used three datasets of triples to build the WKG:
the ``Types'' dataset, which contains subsumption triples of the form (subject,
``is-a,'' Class), where Class is a category of the DBpedia ontology; the
``Properties'' dataset, which contains triples extracted from infoboxes; and the
DBpedia ontology, from which we used all triples with predicate ``subClassOf.''
This last data was used to reconstruct the full ontological hierarchy of the
graph.
We then discarded the predicate part of each triple and conflated all triples
having the same subject and object, obtaining an edge list. In this process, we
discarded all triples whose subject or object belonged to external namespaces
(e.g., FOAF and \url{schema.org}). We also discarded all triples from the
``Properties'' dataset whose object was a date or any other kind of measurement
(e.g., ``Aristotle,'' ``birthYear,'' ``384 B.C.''), because by definition they
never appear as subjects in other triples.

\subsection*{Ideological classification of the US Congress} %
To get a list of ideologies we consider the ``Ideology'' category in the DBpedia
ontology and look up in the WKG all nodes $Y$ connected to it by means of a
statement ($Y$, ``is-a,'' ``Ideology''). We found $M=819$ such nodes (see
\emph{SI appendix} for a complete list of the ideologies). Given a politician
$X$ and an ideology $Y$ we then compute the truth value of the statement ``$X$
endorses ideology $Y$.'' To perform the classification, we use two standard
classifier algorithms: $k$-Nearest Neighbors \cite{Bishop2006} and Random
Forests \cite{Breiman2001}. To assess the classification accuracy we computed
F-score and area under Receiver Operating Characteristic (ROC) curve using
10-fold cross-validation.

\subsection*{Simple factual statements} %
We formed simple statements by combining each of $N$ subject entities with each
of $N$ object entities. We performed this procedure in four subject areas: (1)
Academy Awards for Best Movie ($N = 59$), (2) US presidential couples ($N =
17$), (3) US state capitals ($N = 48$), and (4) world capitals ($N = 187$). For
directors with more than one award, only the first award was used. All data were
taken from Wikipedia (see \emph{SI appendix}). To make the test fair, if a
triple indicating a true statement was already present in the WKG, we removed it
from the graph before computing the truth value. This step of the evaluation
procedure is typical of link prediction algorithms \cite{Liben-Nowell2007}.

\subsection*{Independent corpus of statements} %
The second ground truth dataset is based on the Google Relation Extraction
Corpus (GREC) \cite{Google2013}. For simplicity we focus on two types of
statements, about education degrees ($N = 602$) and institutional affiliations
($N = 10,726$) of people, respectively. Each triple in the GREC comes with truth
ratings by five human raters (Figure~\ref{fig:fig3_relation}(c)), so we map the
ratings into an ordinal scale between $-5$ (all raters replied `No') and $+5$
(all raters replied `Yes'), and compare them to the truth values computed by the
fact checker. The subject entities of several triples in the GREC appear in only
a handful of links in the WKG, limiting the chances that our method can find
more than one path. Therefore we select from the two datasets only triples
having a subject with degree $k > 3$. Similarly to the previous task, if the
statement is already present in the WKG, we remove the corresponding triple
before computing the truth value.

\subsection*{Acknowledgements} 

The authors would like to thank Karissa McKelvey for original exploratory work
and Deborah Rocha for editing the manuscript. We acknowledge the Wikimedia
Foundation, the DBpedia project, and Google for making their respective data
freely available. This work was supported in part by the Swiss National Science
Foundation (fellowship 142353), the Lilly Endowment, the James S. McDonnell
Foundation, NSF (grant CCF-1101743), and DoD (grant W911NF-12-1-0037). The
authors declare that they have no competing financial interests.

\subsection*{Author Contributions} 

GLC, LMR, JB, FM, and AF designed the research. GLC and PS performed simulations
and analyzed the data. GLC, LMR, JB, AF, and FM wrote the manuscript.


\clearpage

\begin{table}
  \caption{\textbf{Transitive closure calibration}. Area under
    ROC curve of two classifiers, random forests (RF) and $k$-nearest neighbors
  ($k$-NN) on the ideological classification task.}
  \centering
  \begin{tabular}{lrcccc}
    \toprule
    & & \multicolumn{2}{c} {Directed} & \multicolumn{2}{c}{Undirected} \\
    & & k-NN & RF & k-NN & RF \\
    \midrule
    \multirow{2}{*}{House}
    & Metric & 96 & 99 &  97 & 99 \\
    & Ultra-metric & 56 & 57 &  53 & 57  \\
    \midrule
    \multirow{2}{*}{Senate}
    & Metric & 70 & 100 & 96 & 100  \\
    & Ultra-metric & 49 & 39 &  70 & 61 \\
    \bottomrule
  \end{tabular}
  \label{tab:calibrationresults}
\end{table}

\begin{table}
    \caption{\textbf{Ideological classification results}. Out-of-sample F-score
    and Area Under Receiver Operating Characteristic (AUROC) of random forest
    (RF) and $k$-nearest neighbors ($k$-NN) classifiers trained on truth scores
    computed by the fact checker, using either the transitive closure or solely
    information from infoboxes.}
    \centering
    \begin{tabular}{ccccc}
        \toprule
        & \multicolumn{2}{c}{RF} & \multicolumn{2}{c}{$k$-NN} \\
        \midrule
        Dataset & F-score & AUROC & F-score & AUROC\\
        \midrule
        \multicolumn{5}{c}{\sc Transitive closure $\mathcal F_\mathrm{tc}$}\\
        \midrule
        Senate & $0.99$ & $1.00$ & $0.91$ & $0.96$\\
        House & $0.99$ & $1.00$ & $0.90$ & $0.97$\\
        \midrule
        \multicolumn{5}{c}{\sc Infoboxes $\mathcal F_\mathrm b$}\\
        \midrule
        Senate & $0.66$ & $0.46$ & $0.62$ & $0.54$\\
        House & $0.54$ & $0.66$ & $0.68$ & $0.54$\\
        \bottomrule
    \end{tabular}
    \label{tab:classification}
\end{table}

\begin{table}
    \caption{\textbf{Agreement between fact checker and human raters}. We use
    rank-order correlation coefficients (Kendall's $\tau$ and Spearman's $\rho$)
    to assess whether the scores are correlated to the ratings. Significance tests
    rule out the null hypothesis that the correlation coefficients are zero.}
    \centering
    \begin{tabular}{lllll}
        \toprule Relation & \multicolumn{1}{c}{$\rho$} &
        \multicolumn{1}{c}{$p$-value} & \multicolumn{1}{c}{$\tau$} &
        \multicolumn{1}{c}{$p$-value} \\
        \midrule
        Degree & $0.17$ & $2\times 10^{-5}$ & $0.13$ & $10\times 1^{-6}$ \\
        Institution & $0.09$ & $4\times 10^{-19}$ & $0.07$ & $1\times 10^{-24}$ \\
        \bottomrule
    \end{tabular}
  \label{tab:relation}
\end{table}

\end{document}